%% file: main.tex
\documentclass[conference, onecolumn]{IEEEtran}
\IEEEoverridecommandlockouts
\usepackage{cite}
\usepackage{amsmath,amssymb,amsfonts}
\usepackage{algorithmic}
\usepackage{graphicx}
\usepackage{textcomp}
\usepackage{algorithm}
\usepackage{algorithmic}
\usepackage[acronym]{glossaries}
\usepackage{dblfloatfix}

\usepackage{orcidlink}
\usepackage{svg}
\newcommand{\orcid}[1]{\href{https://orcid.org/#1}{\includegraphics[width=10pt]{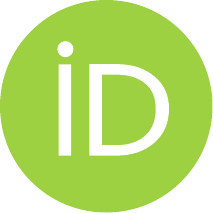}}}

\usepackage{xcolor}
\def\BibTeX{{\rm B\kern-.05em{\sc i\kern-.025em b}\kern-.08em
    T\kern-.1667em\lower.7ex\hbox{E}\kern-.125emX}}
\begin{document}

\title{Forecasting Anonymized Electricity Load Profiles}

\author{\IEEEauthorblockN{Joaquin Delgado Fernandez\orcid{0000-0003-1326-6134}\IEEEauthorrefmark{1}, Sergio Potenciano Menci\orcid{0000-0002-9032-7183}\IEEEauthorrefmark{1}, and Alessio Magitteri\orcid{}\IEEEauthorrefmark{2}}

\IEEEauthorblockA{\IEEEauthorrefmark{1}SnT - Interdisciplinary Center for Security, Reliability and Trust, University of Luxembourg, \\ Luxembourg, Luxembourg, Email: [joaquin.delgadofernandez; sergio.potenciano-menci](at)uni.lu}

\IEEEauthorblockA{\IEEEauthorrefmark{2}Enovos Luxembourg S.A. Esch-sur-Alzette, Luxembourg, Email: [alessio.magitteri](at)enovos.eu}
    
% \thanks{This research was funded in part by the Luxembourg National Research Fund (FNR) and PayPal, PEARL grant reference 13342933/Gilbert Fridgen and by FNR grant reference HPC BRIDGES/2022\_Phase2/17886330/DELPHI. For the purpose of open access and in fulfillment of pen access and fulfilling the obligations arising from the grant agreement, the author has applied a Creative Commons Attribution 4.0 International (CC BY 4.0) license to any Author Accepted Manuscript version arising from this submission. This paper has been supported by Enovos.}

\thanks{This research was funded in part by the Luxembourg National Research Fund (FNR) and PayPal, PEARL grant reference 13342933/Gilbert Fridgen, by FNR grant reference HPC BRIDGES/2022\_Phase2/17886330/DELPHI and by the Luxembourgish Ministry of Economy with grant reference 20230227RDI170010375846. For the purpose of open access and in fulfillment of pen access and fulfilling the obligations arising from the grant agreement, the author has applied a Creative Commons Attribution 4.0 International (CC BY 4.0) license to any Author Accepted Manuscript version arising from this submission. This paper has been supported by Enovos.}
}
% \author{Joaquin Delgado Fernandez\orcid{0000-0003-1326-6134}\IEEEauthorrefmark{1}, Sergio Potenciano Menci\orcid{0000-0002-9032-7183}\IEEEauthorrefmark{1}, and Alessio Magitteri\orcid{}\IEEEauthorrefmark{2}

% \IEEEauthorblockA{\IEEEauthorrefmark{1}SnT - Interdisciplinary Center for Security, Reliability and Trust, University of Luxembourg, \\ Luxembourg, Luxembourg, Email: [joaquin.delgadofernandez; sergio.potenciano-menci](at)uni.lu}

% \IEEEauthorblockA{\IEEEauthorrefmark{2}Enovos Luxembourg S.A. Esch-sur-Alzette, Luxembourg, Email: [alessio.magitteri](at)enovos.eu}

% 

% }   

\maketitle

\begin{abstract}
\input{Sections/00_abstract}

\end{abstract}

\input{Sections/zz_acronym}

\input{Sections/01_introduction}

\input{Sections/03_Methodology}

\input{Sections/04_results}

\input{Sections/05_discussion}

\input{Sections/06_AI}

\bibliographystyle{ieeetr} % or another style such as apalike, ieeetr, etc.
\bibliography{references.bib}

\end{document}

%% file: Sections/00_abstract.tex
In the evolving landscape of data privacy, the anonymization of electric load profiles has become a critical issue, especially with the enforcement of the General Data Protection Regulation (GDPR) in Europe. These electric load profiles, which are essential datasets in the energy industry, are classified as personal behavioral data, necessitating stringent protective measures. This article explores the implications of this classification, the importance of data anonymization, and the potential of forecasting using microaggregated data.
The findings underscore that effective anonymization techniques, such as microaggregation, do not compromise the performance of forecasting models under certain conditions (i.e., forecasting aggregated). In such an aggregated level, microaggregated data maintains high levels of utility, with minimal impact on forecasting accuracy.
The implications for the energy sector are profound, suggesting that privacy-preserving data practices can be integrated into smart metering technology applications without hindering their effectiveness. 


%% file: Sections/zz_acronym.tex
\newacronym{CCPA}{CCPA}{California Consumer Privacy Act}

\newacronym{AI}{AI}{Artificial Intelligence}

\newacronym{DP}{DP}{Differential Privacy}

\newacronym{SecAgg}{SecAgg}{Secure Aggregation}

\newacronym{SMPC}{SMPC}{Secure Multiparty Computation}

\newacronym{IZ}{IT}{Information Technology}

\newacronym{FI}{FI}{Financial Institution}

\newacronym{FedAvg}{Fed-Avg}{Federated Average}

\newacronym{Fed-SGD}{Fed-SGD}{Federated Stochastic gradient descent}

\newacronym{NN}{NN}{Neural Network}

\newacronym{GPU}{GPU}{Graphics Processing Unit}

\newacronym{SME}{SME}{Small and Medium-sized Enterprise}

\newacronym{CSR}{CSR}{Case Study Research}

\newacronym{GT}{GT}{Grounded Theory}

\newacronym{TOE}{TOE}{Technology–Organization–Environment}

\newacronym{EU}{EU}{European Union}

\newacronym{CEP}{CEP}{Clean Energy Package}

\newacronym[plural=ICTs (technologies), longplural=Information and Communication Technologies]{ICT}{ICT}{Information and Communication Technology}

\newacronym{IIoT}{IIoT}{Industrial Internet Of Things}

\newacronym{IoT}{IoT}{Internet Of Things}

\newacronym{DLT}{DLT}{Distributed Ledger Technology}

\newacronym{SGAM}{SGAM}{Smart Grid Architecture Model}

\newacronym{DSO}{DSO}{Distribution System Operator}

\newacronym{TSO}{TSO}{Transmission System Operator}

\newacronym{LFM}{LFM}{Local Flexibility Market}

\newacronym{DER}{DER}{Distributed Energy Resource}

\newacronym{RES}{RES}{Renewable Energy Resources}

\newacronym{EV}{EV}{Electric Vehicle}

\newacronym{DR}{DR}{Demand Response}

\newacronym{HV}{HV}{High Voltage}

\newacronym{MV}{MV}{Medium Voltage}

\newacronym{LV}{LV}{Low Voltage}

\newacronym{SRA}{SRA}{Scalability and Replicability Analysis}

\newacronym{OLTC}{OLTC}{On Load Tap Changer}

\newacronym{DSM}{DSM}{Demand Side Management}

\newacronym{p2p}{p2p}{peer-to-peer}

\newacronym{ETDP}{ETDP}{extended taxonomy design process}

\newacronym{ACER}{ACER}{Agency for the Cooperation of Energy Regulators}

\newacronym{SO}{SO}{system operator}

\newacronym{LEM}{LEM}{local electricity market}

\newacronym{USEF}{USEF}{universal smart energy framework}

\newacronym{FSP}{FSP}{flexibility service provider}

\newacronym{OCPP}{OCPP}{Open Charge Point Protocol}

\newacronym{EFDM}{EFDM}{energy flexibility data model}

\newacronym{CIM}{CIM}{common information model}

\newacronym{GDPR}{GDPR}{General Data Protection Regulation}

\newacronym{NIST}{NIST}{National Institute of Standards and Technology}

\newacronym{OSI}{OSI}{Open Systems Interconnection}

\newacronym{TCP}{TCP}{Transmission Control Protocol/Internet Protocol}

\newacronym{IP}{IP}{Internet Protocol}

\newacronym{UDP}{UDP}{User Datagram Protocol}

\newacronym{IED}{IED}{Intelligent electrical device}

\newacronym{UC}{UC}{use case}

\newacronym{BUC}{BUC}{business use case}

\newacronym{MS}{MS}{Member State}

\newacronym{NECP}{NECP}{National Energy and Climate Plan}

\newacronym{ENTSOE}{ENTSO-E}{European Network of Transmission System Operators for Electricity}

\newacronym{UMEI}{UMEI}{Universal Market Enabling Interface}
 
\newacronym{MO}{MO}{market operator}

\newacronym{CIA}{CIA}{Confidentiality, Integrity and Availability}

\newacronym{EC}{EC}{European Commission}

\newacronym{OMIE}{OMIE}{Iberian Electricity Market Operator}

\newacronym{ENW}{ENWL}{Electricity North West Ltd.}

\newacronym{DPS}{DPS}{Dynamic Procurement System}

\newacronym{ES}{ES}{Spain}

\newacronym{UK}{UK}{United Kingdom}

\newacronym{NL}{NL}{Netherlands}

\newacronym{ITT}{ITT}{Invitation to Tender}

\newacronym{ENA}{ENA}{Energy Networks Association}

\newacronym{NG}{NGED}{National Grid Electricity Distribution}

\newacronym{DNO}{DNO}{Distribution Network Operator}

\newacronym{API}{API}{Application Programme Interface}

\newacronym{ETPA}{ETPA}{Energy Trading Platform Amsterdam}

\newacronym{GMS}{GMS}{Grid and Management Service}

\newacronym{GOPACS}{GOPACS}{Grid Operators Platform for Congestion Spreads}

\newacronym{JSON}{JSON}{JavaScript Object Notation}

\newacronym{IEA}{IEA}{International Energy Agency}

\newacronym{DSR}{DSR}{Design Science Research}

\newacronym{DSRM}{DSRM}{Design Science Research Methodology Process Model}

\newacronym{IS}{IS}{Information Systems}

\newacronym{RP}{RP}{Research Publication}

\newacronym{PV}{PV}{Photovoltaic}

\newacronym{SoS}{SoS}{System of Systems}

\newacronym{FL}{FL}{Federated Learning}

\newacronym{DS}{DS}{Design Science}

\newacronym{SOA}{SOA}{Service-oriented Architecture}

\newacronym{XaaS}{XaaS}{X as a Service}

\newacronym{BRP}{BRP}{Balance Responsible Party}

\newacronym{P2P}{P2P}{Peer-to-Peer}

\newacronym{SaaS}{SaaS}{Software as a Service}

\newacronym{EES}{EES}{Electrochemical Energy Storage}

\newacronym{TLS}{TLS}{Traffic Light System}

\newacronym{UML}{UML}{Unified Modeling Language}

\newacronym{SCADA}{SCADA}{Supervisory Control And Data Acquisition}

\newacronym{ADMS}{ADMS}{Advance Distribution Management System}

\newacronym{RTU}{RTU}{Remote Terminal Unit}

\newacronym{SOTA}{SOTA}{State-Of-the-Art}

\newacronym{ESP}{ESP}{Energy Synchronization Platform}

\newacronym{CP}{CP}{Company Platform}

\newacronym{MP}{MP}{Market Platform}

\newacronym{TF}{TF}{Task Force}

\newacronym{PF}{PF}{Power Flow}

\newacronym{OPF}{OPF}{Optimal Power Flow}

\newacronym{SSU}{SSU}{Smart Storage Unit}

\newacronym{MPOPF}{MPOPF}{Multi-Period Optimal Power Flow}

\newacronym{VPP}{VPP}{Virtual Power Plant}

\newacronym{HEMS}{HEMS}{Home Energy Management System}

\newacronym{ESS}{ESS}{Energy Storage System}

\newacronym{ML}{ML}{Machine Learning}

\newacronym{DL}{DL}{Deep Learning}

\newacronym{STLF}{STLF}{Short-Term Load Forecasting}

\newacronym{LSTM}{LSTM}{Long Short-Term Memory}

\newacronym{DTW}{DTW}{Dynamic Time Warping}

%% Copy the following one to add more acronyms
% \newacronym{}{}{}

\newacronym{USA}{USA}{United States of America}

\newacronym{NILM}{NILM}{Non-intrusive load monitoring}

\newacronym{FCL}{FCL}{Fully Connected Layers}

\newacronym{RMSE}{RMSE}{Root Mean Square Error}

\newacronym{NRMSE}{NRMSE}{Normalized Root Mean Square Error}

\newacronym{MAPE}{MAPE}{Mean Absolute Percentage Error}

\newacronym{MAE}{MAE}{Mean Absolute Error}

\newacronym{MSE}{MSE}{Mean Squared Error}

\newacronym{MASE}{MASE}{Mean Absolute Scaled Error}

\newacronym{DBI}{DBI}{Davies–-Bouldin index}

\newacronym{CPO}{CPO}{Charging Point Operator}

\newacronym{CS}{CS}{Charging Station}

\newacronym{EE}{EE}{Energy Efficiency}

\newacronym{NREL}{NREL}{National Renewable Energy Laboratory}

\newacronym{CNN}{CNN}{Convolutional Neural Network}

\newacronym{EVCP}{EVCP}{electric vehicle Charging Pile}

\newacronym{CPU}{CPU}{Central Processing Unit}

\newacronym{PID}{PID}{proportional-integral-derivative}

\newacronym{SVM}{SVM}{suppport vector machine}

%% file: Sections/01_introduction.tex
\section{Introduction}

In early 2009, the European Commission, through Directive 2009/72/EC, mandated the widespread deployment of smart meters across Europe~\cite{eu_sm}. These devices are key in the digitalization of the energy sector~\cite{gungor2011smart}, allowing automatic granular recording of energy consumption, covering electricity, natural gas, district heating and water use. In the electricity sector, smart meters typically record consumption data at a 15-minute resolution, generating what are known as load profiles. These devices also facilitate remote data transmission to the designated \gls{DSO}, which subsequently shares the data with energy suppliers. Smart meters load profiles play a critical role for \glspl{DSO} and energy suppliers, as they are essential for tasks such as congestion management, peak load analysis, and consumption forecasting~\cite{petropoulos2022forecasting}. For energy suppliers, these profiles have become particularly valuable for understanding the evolving consumption behaviors of customers, driven by the increasing integration of \gls{RES}, such as photovoltaic systems and batteries, as well as the adoption of new electric technologies like \gls{EV}.

Despite the clear advantages of smart meters—such as automated, remote data collection and enhanced granularity~\cite{6520030}—their use faces significant challenges within the European Union. One major constraint arises from the \gls{GDPR}~\cite{gdpr_eu}, which has been in force since May 25, 2018. Under the \gls{GDPR}, explicit user consent is required to process load profile data. Another constraint stems from the high granularity of smart meter data. As the resolution increases from hourly values to 15-minute intervals, 5-minute intervals, or even real-time data, it becomes possible to infer sensitive details about consumers’ lifestyles, daily routines, and home occupancy patterns. These combined challenges make it difficult to utilize smart meter data for third-party applications or even for energy suppliers, as \gls{GDPR} imposes stringent measures to safeguard personal data, and consumers may hesitate to consent to the use of their granular load profiles for tasks like forecasting.

One potential solution to these constraints is pseudo-anonymization, which involves removing personal and quasi-identifiable information from smart meter data~\cite{efthymiou2010smart}. Since pseudonymization preserves consumer profiles intact, information regarding the consumers can still be accessed, presenting notable challenges to privacy. In summary, fully anonymizing data presents difficulties since energy consumption is intrinsically time-dependent, and mere removal of identifiers does not guarantee data anonymization.

The nature of time series in which the values at time $t$ are closely related to the ones at $t-1$ and $t+1$ reflects sequential and temporal dependencies. This characteristic limits the effectiveness of classical anonymization techniques based on data perturbations or shuffling, leaving non-perturbative techniques as viable alternatives~\cite{domingo2005ordinal}.

Non-perturbative anonymization methods, however, come with trade-offs. As the level of anonymization increases, the utility of the data typically diminishes~\cite{WANG2023100132}. Reduced data utility can adversely impact forecasting capabilities, making it more challenging for energy suppliers to balance supply and demand or to accurately predict consumption patterns.

In this paper, we examine the impact of a specific non-perturbative anonymization method, microaggregation, on the accuracy of energy demand forecasting, particularly for the \gls{STLF} with a one hour horizon. Microaggregation groups similar data points to obscure individual identities while preserving key dataset properties. Our analysis focuses on assessing how microaggregation affects the accuracy of energy demand forecasts. Results indicate that microaggregated data has minimal impact on forecasting accuracy at an aggregated level. This finding is significant, as it demonstrates that high levels of data utility can be maintained while ensuring user privacy. It allows energy suppliers and third parties, for instance, under project or service agreements, to work together to utilize smart meter data without direct user consent while remaining compliant with \gls{GDPR}, safeguarding individual privacy, and retaining data utility.

%% file: Sections/03_Methodology.tex
\section{Background}
 \subsection{Dataset}
 \label{subsec:data}
In our study, we used a dataset from the Low Carbon London project, conducted by UK Power Networks between November 2011 and February 2014 in the London area \cite{strbac_low_2024}. This dataset, which records household electrical consumption in kilowatt-hours (kWh) at half-hour intervals, served as the foundation for our simulations.

The original dataset comprised more than 4000 households with electric measurements in half-hour resolution from 1st of January 2013 until 31st of December 2013. We selected a random sample of 1000 households from the original dataset to limit the computational requirements of the experiments.

\subsection{Anonymization}

In the context of anonymizing electric load data, the focus resides on time series, specifically the measurement time (when) and the value (what). Other features, such as name, surname, Id, network, and additional attributes, are excluded from consideration. This results in a simplified dual tuple (when, what), where each element alone is insufficient for identification. However, when these tuples are analyzed as a time series, they can still enable the identification of consumers. Consequently, this data remains personal information and necessitates anonymization for third-party use.

There are various techniques for data anonymization, each offering different levels of robustness. These techniques fall into two main categories: generalization, which does not alter the data but reduces detail, and perturbation which modifies the data by replacing it with other values.

Given the temporal nature and the limitations it imposes, like not distorting time-dependent patterns or applying noise that could affect forecasting. We restricted ourselves to generalization methods such as k-anonymity and microaggregation.

K-anonymity\cite{sweeney2002k}, a technique within generalization, aims to prevent a data subject from being singled out by grouping them with at least $k$ other individuals. This is achieved by generalizing attribute values so that each individual shares the same value. For example, individual households’ consumption values are aggregated with those of $k$ other individuals and averaged, making the output indistinguishable from the outside.

In a different manner, in 2005, Jossep Domingo-Ferrer and Vicenç Torra \cite{domingo2005ordinal} published an evolution to K-anonymity named Microaggregation. Their algorithm create based on simulart and not randomly groups of size $k$. Their algorithm named Maximum Distance to Average Vector (MDAV) is as follows:

% An evolution of this was proposed by \cite{domingo2005ordinal} in which the groups of size $k$ are selected by similarity and not randomly. This newer approach increases the utility of the groups. 

% In 2005, Jossep Domingo-Ferrer and Vicenç Torra \cite{domingo2005ordinal} published an evolution to K-anonymity named Microaggregation in which they proposed solutions to tackle the loss in utility of k-anonymity by using intelligent aggregation of time series. Their algorithm named Maximum Distance to Average Vector (MDAV) is as follows:

\begin{algorithm}
\caption{MDAV (R: dataset, k: integer) from \cite{domingo2005ordinal}}
\begin{algorithmic}[1]
    \WHILE{$|R| \geq 3k$}
        \STATE Compute the average record $\hat{x}$ of all records in $R$. 
        \STATE Consider the most distant record $x_r$ to the average record $\hat{x}$ using a distance metric
        \STATE Find the most distant record $x_s$ from the record $x_r$.
        \STATE Form two clusters around $x_r$ and $x_s$. One cluster contains $x_r$ and the $k-1$ records closest to $x_r$. The other cluster contains $x_s$ and the $k-1$ records closest to $x_s$.
        \STATE Consider as a new dataset $R$ the previous dataset $R$ without the clusters: $x_r$ and $x_s$ 
    \ENDWHILE
    \IF{$3k-1 \leq |R| < 2k$}
        \STATE Compute the average record $\hat{x}$ of the remaining dataset $R$.
        \STATE Find the most distant record $x_r$ from $\hat{x}$.
        \STATE Form a cluster containing $x_r$ and the $k-1$ records closest to $x_r$.
        \STATE Form another cluster containing the rest of the records.
    \ENDIF
\end{algorithmic}
\end{algorithm}

Recent research underscores that microaggregation alone may not be sufficient to fully protect smart grid data from disclosure risks. To address this, the DFTMicroagg algorithm was proposed, offering a dual-level perturbation mechanism to enhance privacy in energy systems \cite{adewole2022dftmicroagg}

The algorithm leverages the benefits of Discrete Fourier Transform (DFT) and microaggregation to provide an additional layer of protection.
The DFTMicroagg algorithm first applies a low-pass filtering step using DFT to anonymize the data before applying the MDAV microaggregation. 

Although the DFTMicroagg algorithm offers enhanced privacy protection by combining DFT with microaggregation, we decided not to use it due to its complexity and potential impact on data utility. The dual-level perturbation could introduce distortions that affect the accuracy of forecasting models. 

To this end, we focused on simpler generalization methods like MDAV to achieve a balance between privacy and utility.

\begin{figure}[h!]
\centering
        \includegraphics[width=\columnwidth]{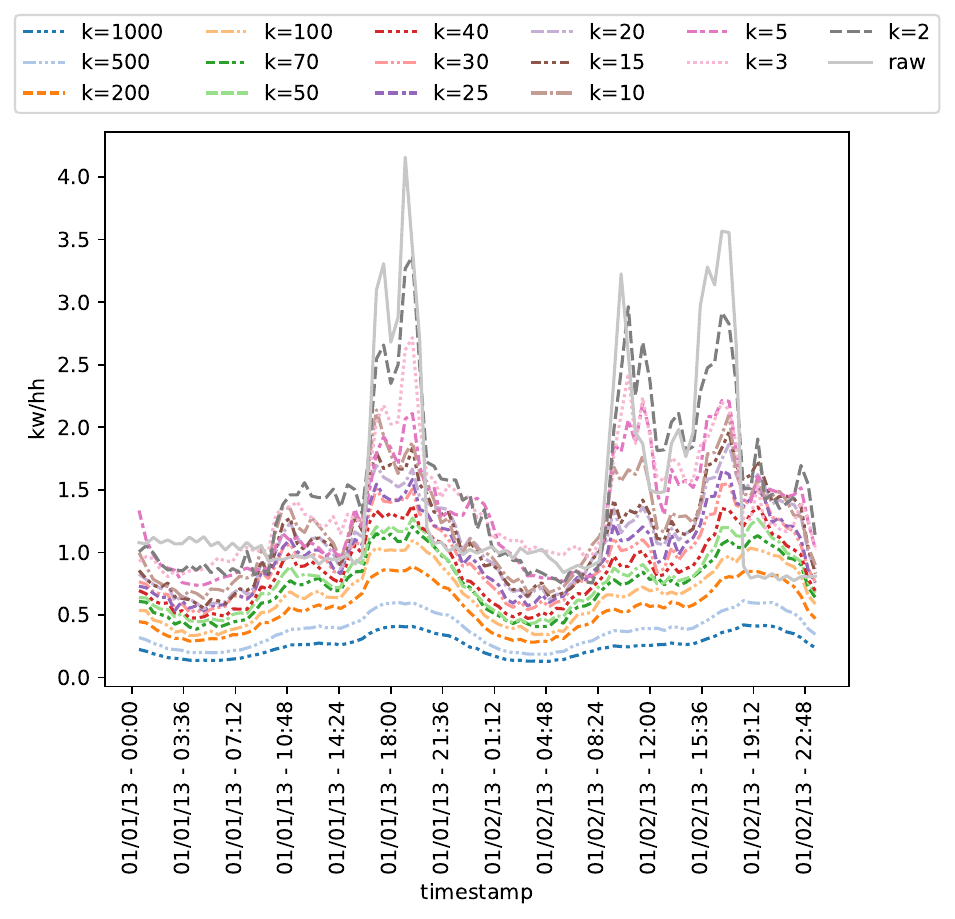}
    \caption{Visual representation of the different datasets under different anonymization levels.}
    \label{fig:anom_ks}
\end{figure}

Figure\ref{fig:anom_ks} represents one day of consumption for a given household and the groups the household is part of under different levels of anonymization.

\subsection{Forecasting}
To forecast household electrical consumption and evaluate model performance under varying privacy levels, we compared different models, which we split into four groups:  1) baseline approaches, 2) statistical models, 3) machine learning (ML) models, and 4) neural network (NN) architectures. The baseline approach served as a reference point for evaluating other models. For statistical models, we selected MSTL\cite{bandara2021mstl} and MFLES\cite{mfles}. Our ML models included XGBoost\cite{xgboost} and LGBM\cite{lightgbm}. Lastly, we employed six NN architectures divided in two subcategories. Firstly the MLP-based models~\footnote{Referred as NN from now on in the paper.}: MLP, NBEATS\cite{nbeats} and NHITS\cite{nhits} and the transformer-based models: Autoformer\cite{autoformer}, Informer\cite{informer}, TFT\cite{tft}.

Each category of models was configured differently. For the statistical models, we used the statsforecast library, setting the seasonal length to 48 as the data is in half-hour resolution, which corresponds to one day in half-hour resolution. Specifically, for MSTL, we incorporated AutoARIMA as it requires a seasonal forecaster.

For the ML models, we utilized the mlforecast library. We accounted for seasonality by including lags of 1, 48, and 336 being those previous half-hour, previous day and previous week. These lags were further transformed using Expanding Mean and Rolling Mean with a window size of 24. We also differentiated the target values with lags of 48 and 336, representing one day and one week, respectively. Additionally, we included the day of the week as an exogenous feature.

For the NN models, we employed the neuralforecast library. We set the local scaler to be robust to outliers using a robust local scaler. To manage computational demands, we empirically set the input size to 500 values.

This structured approach allowed us to systematically evaluate the trade-offs between data privacy and forecasting accuracy across different modeling techniques.

\subsection{Simulation Environment}
We performed the simulations in the IRIS Cluster of the high-performance computer (HPC) facilities of the University of Luxembourg~\cite{HPC}. The simulations for the microaggregation ran in a specific node with 1 Tb of RAM while the forecasting models trained on one NVIDIA Tesla V100 with 16 Gb or 32 Gb depending on the allocation. We programmed the forecasting using the Nixtla stack \cite{olivares2022library_neuralforecast,garza2022statsforecast}.

\subsection{Simulation Procedure}
To evaluate the impact of $k$ on forecasting, we generated anonymized data sets $n$ by applying micro-aggregation to the original dataset. We defined nine levels of privacy as: $k=2$, $k=3$,  $k=5$,  $k=10$,  $k=15$,  $k=20$,  $k=25$,  $k=30$,  $k=40$, $k=50$,  $k=70$,  $k=100$, $k=200$, $k=500$, $k=1000$. 

With this, we created a set of different datasets with 500 the least private case ($k=2$) to 1 single group in the most private scenario with $k=1000$ as an extreme case. This last group is equivalent to averaging all the profiles into a unique time series. In the opposite scenario, we keep the data without anonymization under the label \textit{raw} in the upcoming experiments.

\begin{figure*}[b!]
\centering
        \includegraphics[width=\textwidth]{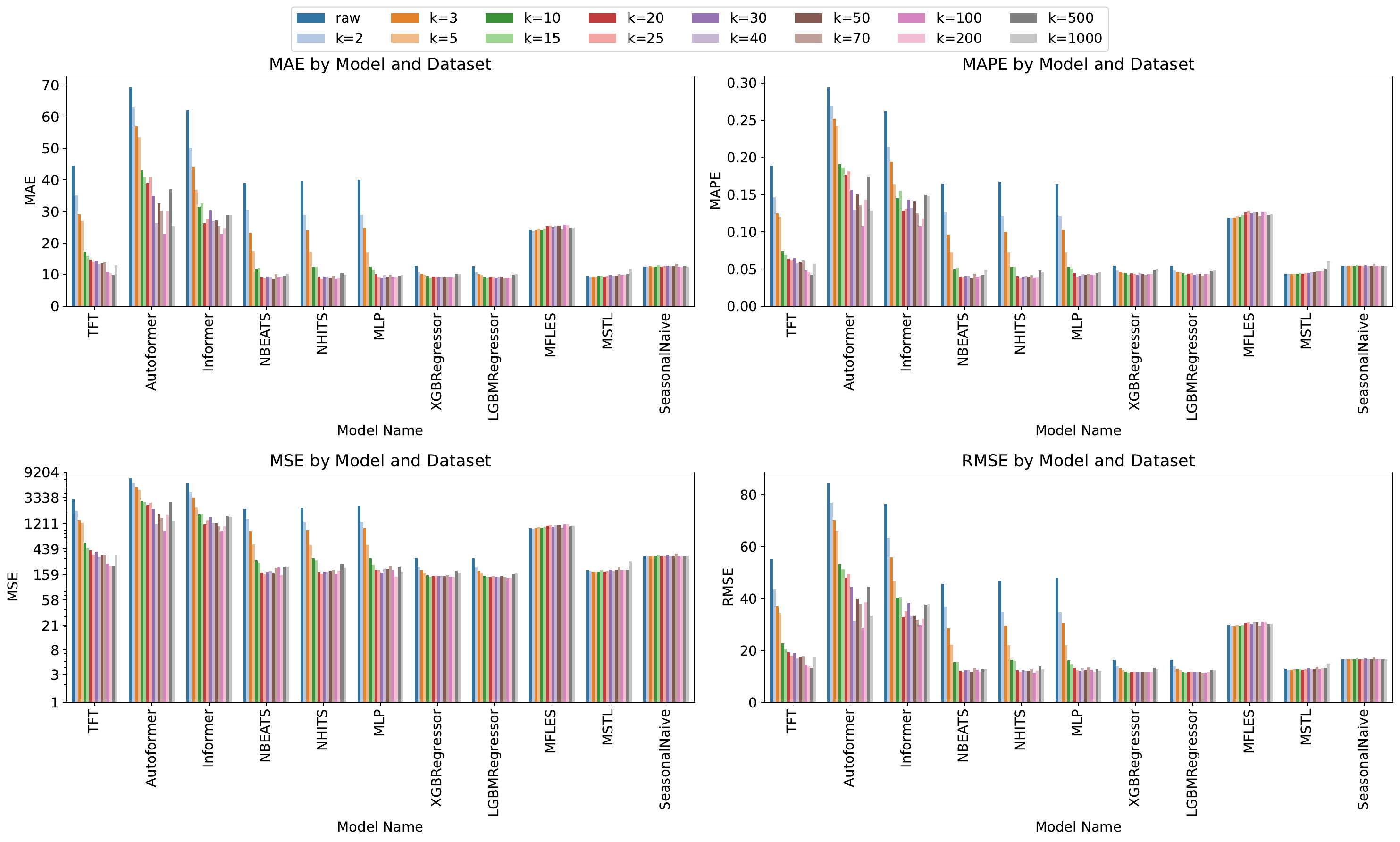}
    \caption{Overall performance results by $k$ anonymization level in terms of mean ans standard deviation.}
    \label{fig:results}
\end{figure*}

To evaluate the forecasting performance at different times in the year on the anonymized datasets and non-anonymized data, we employed a cross-validation approach. To minimize computational load, the forecasting process was divided into 5 stages or windows: (1) training on data up to August 28th and forecasting for August 29th, (2) training on data up to September 28th and forecasting for September 29th, (3) training on data up to October 29th and forecasting for October 30th. In the (4) stage, we train up to the 29th of November and forecast on the 30th, and lastly in (5) we train up to the 30th of December and forecast on the 31st. For each of the windows, the model retrains again to account for the new data.  
We then produce day-ahead forecasts; meaning we forecast 24 hours in advance. Since the dataset is in half-hour resolution, we forecast 48 values.

Once individual households or groups are forecasted, we aggregated them; in the raw case by simply summing across the households for a given timestamp; see Equation \eqref{eq:agg_raw} and on the anonymized sets by first multiplying by the anonymization factor $k$ and then summing across by timestamp as in Equation \eqref{eq:agg_anom}.
\begin{equation}
\label{eq:agg_raw}
    \text{Aggregated\_Raw}(t) = \sum_{i=1}^{n} \text{Household}_i(t)
\end{equation}

\begin{equation}
\label{eq:agg_anom}
\text{Aggregated\_Anonymized}(t) = \sum_{i=1}^{\frac{n}{k}} (k \cdot \text{group}_i(t))
\end{equation}

We compared the performance of these aggregated profiles over the 5 days using various evaluation metrics, including Mean Absolute Error (MAE), Mean Absolute Percentage Error (MAPE), Mean Squared Error (MSE), Root Mean Squared Error (RMSE), and Symmetric Mean Absolute Percentage Error (SMAPE). To account for variability and the stochastic nature of the models, we repeated the analysis twice.

%% file: Sections/04_results.tex
\section{Results}

The evaluation of various forecasting methods is classified into four distinct groups: Baseline, Statistical, Machine Learning, and Neural Networks. These models are trained at varying levels of microaggregation, represented by $k$. Their performance is evaluated using multiple metrics, aiming to offer a comprehensive analysis of how these techniques function across different data aggregation levels. The results obtained we display in \autoref{fig:results}.

% \begin{figure*}[b!]
% \centering
%         \includegraphics[width=\textwidth]{figures/result.pdf}
%     \caption{Overall performance results by $k$ anonymization level in terms of mean ans standard deviation.}
%     \label{fig:results}
% \end{figure*}

\subsection{Comparison Across Categories}

The baseline approach, represented by \texttt{Seasonal Naive}, overall exhibits constant metrics at all levels of microaggregation. This indicates that although simple and computationally inexpensive, the baseline approach can forecast accurately and is a valid baseline for the rest of the models.

Statistical approaches, including \texttt{MSTL} and \texttt{MFLES}, displayed mixed performance over the baseline. Specially, \texttt{MSTL} shows relatively stable performance across different $k$ levels. In contrast, \texttt{MFLES} has more variability in terms of performance, being worse than the baseline across all $k$ levels. This indicates that while \texttt{MSTL} can maintain consistent accuracy despite increasing anonymization, \texttt{MFLES} is more sensitive to these changes.

ML approaches, such as \texttt{XGBRegressor} and \texttt{LGBMRegressor}, demonstrate competitive performance, often achieving lower error metrics compared to statistical methods. For instance \texttt{LGBMRegressor} with $k=100$ obtained the best overall MSE and RMSE with values of 138.38 and 11.43 respectively.

These approaches leverage ensemble techniques and gradient boosting, which enhance their ability to model complex relationships within the data. Furthermore, ML approaches handle changes in $k$ relatively well, maintaining competitive and constant performance across different levels. 

Within the NN approaches, two subfamilies perform differently. On the one hand the MLP-base models namely: \texttt{NHITS}, \texttt{MLP}, \texttt{NBEATS} displays a good performance in general. Particularly performant for \texttt{NBEATS} with $k=50$ for which it obtained the overall best MAE, MAPE and SMAPE of 8.65, 0.037 and 0.018 respectively.
On the other hand the transformer-based models notably: \texttt{Autoformer}, \texttt{Informer} and \texttt{TFT} achieve the overall worst performance across most metrics. In particular the \texttt{Autoformer} when trained on raw data is the overall worst-performing model of the evaluation with 69.39, 0.29, 7286.95, 84.43 and 0.16 of MAE, MAPE, MSE, RMSE and SMAPE respectively. These results suggest a severe overfitting case.  

In general for both subfamilies we can oversee a similar pattern in which the model performance increases as the $k$ increases. It is noteworthy to mention that under high anonymization regimes the models found the best performing stage. Particularly the DL models find their optimal $k = 60 \ (\sigma = 36)$, similarly the ML models find their minima at slightly less restrictive anonymity  $k = 45 \ (\sigma = 38)$. On the contrary Transformers find their best performing regime at  $k = 233 \ (\sigma = 230)$,

\texttt{NBEATS} is an exemplary model with an accentuated performance increase as k increases. \texttt{NBEATS} performs suboptimal on raw data but has a considerable improvement on high anonymization levels. For instance, the MAE of \texttt{NBEATS} ranges from 38.92 kw/hh using raw data to 8.65 kw/hh when trained with ($k=50$) when it finds their performance maxima.

% \begin{figure}[h!]
% \centering
%         \includegraphics[width=\columnwidth]{figures/nbeats.pdf}
%     \caption{NBEATS performance for each of the $k$ levels}
%     \label{fig:nbeats}
% \end{figure}

\subsection{Impact of Microaggregation Levels}
\label{subsec:micro_impact}

% \begin{equation}
% \label{eq:IL}
%     IL(X, \hat{X}) = \frac{1}{TN} \sum_{j=1}^{T} \sum_{i=1}^{N} \frac{|x_{ij} - \hat{x}_{ij}|}{\sqrt{2} \sigma_j}
% \end{equation}

% \begin{equation}
% \label{eq:SSE}
% \text{SSE} = \sum_{i=1}^{N} \sum_{j=1}^{M} (X_{ij} - \hat{X}_{ij})^2 
% \end{equation}

% \begin{equation}
% \label{eq:volatility}
% Volatility = \sqrt{\frac{1}{N-1} \sum_{i=2}^{n} \left( \frac{\hat{X}_{i} - \hat{X}_{i-1}}{\hat{X}_{i-1}} - \mu \right)^2} 
% \end{equation}

As the microaggregation level increases; changes in the dynamics of the dataset and thus the way models learn. To evaluate the impact of microaggregation on the datasets, we used the Sum of Squared Errors (SSE)~\cite{domingo2008privacy} (see Eq.~\eqref{eq:SSE}) and the Information Loss (IL) ~\cite{yancey2002disclosure} (see Eq.~\eqref{eq:IL}) to measure the information loss and finally to analyze how the dataset's volatility changes as we increase in anonymization level we used Eq.~\eqref{eq:volatility} In the previous equations, $\hat{X}$ represents the anonymized dataset, while $X$ corresponds to the original dataset. Here, $N$ denotes the number of elements to be anonymized, and $M$ indicates the number of observations per element.

\begin{equation}
\label{eq:SSE}
\text{SSE} = \sum_{i=1}^{N} \sum_{j=1}^{M} (X_{ij} - \hat{X}_{ij})^2 
\end{equation}

\begin{equation}
\label{eq:IL}
    IL(X, \hat{X}) = \frac{1}{TN} \sum_{j=1}^{T} \sum_{i=1}^{N} \frac{|x_{ij} - \hat{x}_{ij}|}{\sqrt{2} \sigma_j}
\end{equation}

\begin{equation}
\label{eq:volatility}
Volatility = \sqrt{\frac{1}{N-1} \sum_{i=2}^{n} \left( \frac{\hat{X}_{i} - \hat{X}_{i-1}}{\hat{X}_{i-1}} - \mu \right)^2} 
\end{equation}
To ensure a more robust utility-volatility analysis, we generated 10 distinct microaggregated datasets with 1000 households each from our original set in \ref{subsec:data}. The bold line represents the average, and the shaded area represents the standard deviation of the runs. The figure illustrates that there are no substantial variations among the datasets analyzed.

% In Figure~\ref{fig:vol_loss}, the volatility in the dataset decreases as $k$ increases. To illustrate this loss, we fitted an exponential decay function, $$f = 1.62 \cdot e^{-\frac{t}{3.68}}$$, which fits with an $r^2$ value of 0.86.

\begin{figure}[h!]
\centering
        \includegraphics[width=\columnwidth]{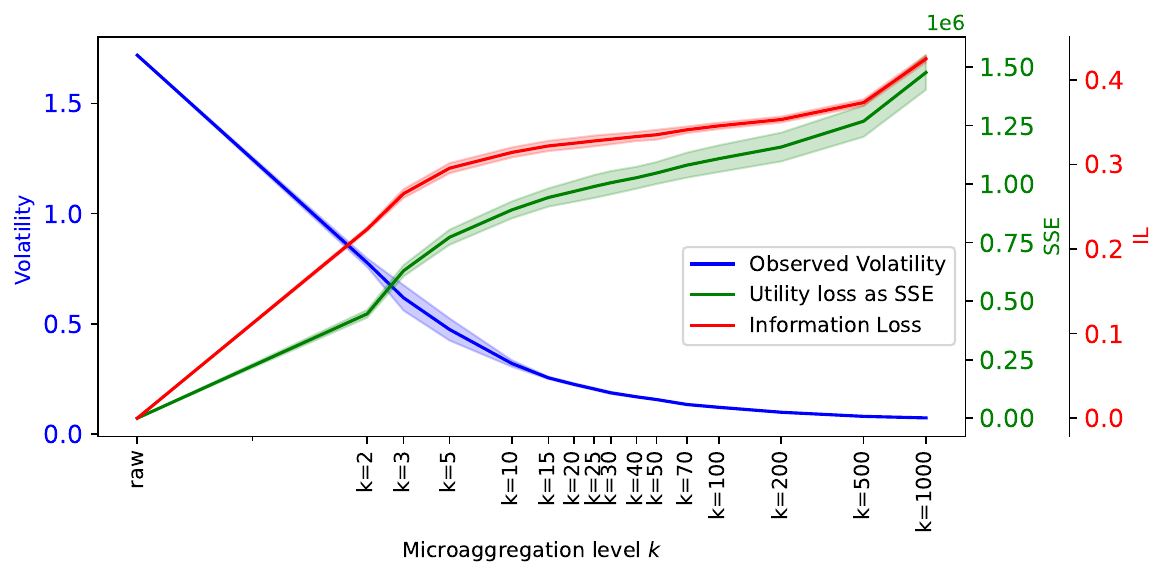}
    \caption{Volatility and information loss for each of the $k$ levels in logarithmic scale.}
    \label{fig:vol_loss}
\end{figure}

In Figure~\ref{fig:vol_loss}, the volatility in the dataset decreases as $k$ increases. To illustrate this loss, we fitted an exponential decay function (see Eq.~\eqref{eq:expo}), which fits with an $r^2$ value of 0.86.

\begin{equation}
\label{eq:expo}
 f = 1.62 \cdot e^{-\frac{t}{3.68}}
 \end{equation}

The quasi-exponential decay in variability within the time series may explain why more complex and larger models tend to overfit the noise, whereas simpler, more trend-focused models perform better in highly volatile scenarios. Overall, as $k$ increases, the performance of the models also improves. A similar pattern is observed in clustering because the data points within each cluster become more homogeneous. This reduction in within-cluster variability makes it easier for forecasting models to identify patterns and forecast accurately \cite{fernandez2023powertech}.

Furthermore, the SSE quantifies the total deviation of data points from their cluster centroids after microaggregation and IL measures the average difference between the actual values and their approximations, adjusted by the variability of each feature in the data.

As $k$ increases, the number of data points in each group also increases, leading to a reduction in the group volatility. This is because larger clusters tend to average out more of the individual variations, resulting in more homogeneous centroids. This information is also visible as the decay in variability indicates that the clusters are becoming more uniform, which simplifies the structure of the data.

Consequently, simpler models that focus on broader trends rather than intricate details perform better, as they are less likely to overfit the noise present in the data. Overall: models, IL and SSE and volatility stabilize at around $k=15$, as indicated by the red dotted line. This statement poses profound implications for both researchers and practitioners. Our experiments display that after $k=15$, adding anonymization to the model is basically a free lunch. Before this point, anonymization is pricey since it produces a quasi-exponential increase in information loss.

%% file: Sections/05_discussion.tex
\section{Discussion}
The findings of this study underscore the viability of using microaggregation as an effective anonymization technique for electric load profiles without significantly compromising the accuracy of forecasting models by evaluating them at an aggregated level. This is particularly relevant in the context of the GDPR, which mandates stringent measures to protect personal data. By demonstrating that microaggregated data can maintain high levels of utility, this research addresses a critical challenge in the energy sector: balancing privacy with data usability.

Our findings indicate that, for instance, \glspl{DSO} or energy suppliers could provide access to their data set by anonymizing their data to comply with regulations. Yet, at the same time, energy suppliers could use such an approach at an aggregated level because it provides clustering benefits, effectively killing two birds with one stone. The reduced volatility within groups enhances model learning by minimizing noise, which improves model fit as group size increases. Additionally, our results suggest a \textit{plateau of information loss} between $k=15$ to $k=200$, where the increase in SSE and IL is not as linear as the rise in protection levels, suggesting an optimal range for balancing privacy and data utility.

% The implications of these findings are significant for energy suppliers and policymakers. Effective anonymization techniques like microaggregation can facilitate the widespread adoption of smart metering technologies by alleviating privacy concerns. This, in turn, can enhance the efficiency and resilience of energy systems through improved consumption forecasting and management.

However, this study is not without limitations. The public dataset used, while extensive, is limited to a specific geographic region, time frame, and year and might not capture current trends (decentralization and electrification of assets). Additionally, while microaggregation has potential, further exploration of other anonymization techniques such as \cite{adewole2022dftmicroagg} and their impact on forecasting accuracy is needed over larger periods to seek consistency. Lastly, this paper evaluates the forecasting performance at an aggregated level; further research could evaluate to what extent this modelization could be applied to household-level forecasting. 

\section{Conclusion}
In conclusion, this study suggests that microaggregation is a viable technique for anonymizing electric load profiles, effectively balancing the need for privacy with the utility of the data. The minimal impact on forecasting accuracy observed across various models, in the case of aggregated level data, underscores the potential for integrating such anonymization methods into the energy sector. Such a technique could help unblock the scarcity and data complexity acquisition of smart meter data for research institutions, universities, or third-party companies while still complying with current European regulations and focusing on maintaining consumer privacy.  

% In conclusion, this study suggests that microaggregation is a viable technique for anonymizing electric load profiles, effectively balancing the need for privacy with the utility of the data. The minimal impact on forecasting accuracy observed across various models underscores the potential for integrating such anonymization methods into the energy sector. This research contributes to the ongoing discourse on data privacy and utility, offering practical insights for energy suppliers and policymakers.

% The significance of these findings lies in their potential to drive the adoption of smart metering technologies, thereby enhancing the efficiency and sustainability of energy systems. 

%% file: Sections/06_AI.tex
\section*{Declaration of Generative AI and AI-assisted technologies in the writing process}
During the preparation of this work, the author used DeepL~\cite{DeepL} and ChatGPT~\cite{ChatGPT} to paraphrase and fix grammatical mistakes. After using this tool/service, the author reviewed and edited the content as needed and take full responsibility for the content of the publication.